\documentclass[12pt]{article}
\usepackage{epsfig}
\usepackage{color}
\usepackage{amssymb,amsmath}
\usepackage{graphicx}
\usepackage{here}
\usepackage{epsfig}

\newcommand{\ba}{\begin{align}}
\newcommand{\ea}{\end{align}}

 \makeatletter
\def\alt{\mathrel{\mathpalette\gl@align<}}
\def\agt{\mathrel{\mathpalette\gl@align>}}
\def\gl@align#1#2{\lower.6ex\vbox{\baselineskip\z@skip\lineskip\z@
\ialign{$\m@th#1\hfil##\hfil$\crcr#2\crcr\sim\crcr}}} \makeatother

\begin{document}
\begin{flushright}
\end{flushright}
\vspace*{1.0cm}

\begin{center}
\baselineskip 20pt 
{\Large\bf 
Multiple-point principle realized with strong dynamics
}
\vspace{1cm}

{\large 
Naoyuki Haba \ and \ Toshifumi Yamada
} \vspace{.5cm}

{\baselineskip 20pt \it
Graduate School of Science and Engineering, Shimane University, Matsue 690-8504, Japan
}

\vspace{.5cm}

\vspace{1.5cm} {\bf Abstract} \end{center}

We present a novel extension of the Standard Model which fulfills the multiple-point principle
 without contradicting the Higgs particle mass measurement.
In the model, the scalar potential has two minima where the scalar field has vacuum expectation values of 246~GeV and the Planck mass~$\simeq 2.44\times 10^{18}$~GeV,
 the latter of which is realized by considering a classically scale invariant setup and requiring that the scalar quartic coupling and its beta function vanish at the Planck scale.
The Standard Model Higgs field is a mixture of an elementary scalar and composite scalars in a new strongly-coupled gauge theory,
 and the strong dynamics gives rise to the negative mass for the SM Higgs field,
 and at the same time, causes separation of the SM Higgs quartic coupling and the quartic coupling for the elementary scalar,
 which leads to the vanshing of the latter quartic coupling and its beta function at the Planck scale.
The model predicts new scalar particles with about 300~GeV mass possessing electroweak charges and Yukawa-type couplings with Standard Model fermions,
 and a new light gauge boson that couples to Standard Model fermions.

\thispagestyle{empty}

\newpage

\setcounter{footnote}{0}
\baselineskip 18pt
%

The multiple-point principle (MPP)~\cite{mpp} is a conjecture that Nature fixes fundamental parameters in such a way that multiple degenerate vacua coexist.
It is further assumed that there are at least two vacua in the Standard Model (SM) scalar potential, one corresponding to our electroweak symmetry breaking vacuum and 
 another to a vacuum where the Higgs field takes a vacuum expectation value (VEV) at the Planck scale $M_P \simeq 2.44\times10^{18}$~GeV.
The MPP demands that the effective potential for the Higgs field whose coupling constants are renormalization-group (RG)-improved, $V(\phi)=-\frac{1}{2}m^2(\phi) \phi^2 + \lambda(\phi) \phi^4$,
 satisfy at the Planck scale
\begin{align}
m^2(\phi\simeq M_P) &= 0, \ \ \ \ \ \lambda(\phi\simeq M_P) = 0, \ \ \ \ \ \beta_\lambda(\phi\simeq M_P) \equiv \frac{{\rm d}}{{\rm d}\log \phi}\lambda(\phi)\vert_{\phi\simeq M_P} = 0.
\label{mpp}
\end{align}
These can be fulfilled if the theory is classically scale invariant~\cite{csi} at the Planck scale and if the Higgs quartic coupling and its beta function simultaneously vanish at that scale.
Unfortunately, the recent precise Higgs particle mass measurement reporting $m_h=125.09\pm 0.24$~GeV~\cite{combinedhiggs},
 combined with analysis of the RG evolutions of the Higgs quartic coupling and other SM couplings~\cite{rge,rgeanalyses}, suggests that it is unlikely to have
 $\lambda(\phi\simeq M_P)=\beta_\lambda(\phi\simeq M_P)=0$ in the SM and its extensions with classical scale invariance.
\footnote{
If one relaxes the requirement of classical scale invariance, it is possible to have $\lambda(\phi\simeq M_P)=\beta_\lambda(\phi\simeq M_P)=0$ in a simple extension of the SM. See Ref.~\cite{mppwithmass}.
}

In this paper, we propose a novel extension of the SM which satisfies the conditions of Eq.~(\ref{mpp}) and is consistent with the measured Higgs particle mass.
Our model is classically scale invariant at high energy scales and the SM Higgs field mass term is generated dynamically in a new strongly-coupled gauge theory.
A salient feature of the model is that the SM Higgs field is a mixture of an elementary scalar field and composite scalar fields in the new gauge theory.
The SM Higgs quartic coupling, $\lambda^{{\rm SM}}$, results from the quartic coupling for the elementary scalar field, $\lambda$, which are related as
 $\lambda^{{\rm SM}} = c_H^4 \, \lambda$, with $0< c_H < 1$ being the fraction of the elementary scalar in the SM Higgs field.
Hence, the quartic coupling for the elementary scalar is enhanced by $1/c_H^4$ compared to that for the SM Higgs field and can attain $\lambda(\phi\simeq M_P)=\beta_\lambda(\phi\simeq M_P)=0$ with an appropriate choice of $c_H$.
The mixing of the elementary and composite scalars is also responsible for dynamical generation of the negative Higgs field mass by the bosonic seesaw mechanism~\cite{bosonicseesaw},
 which is also utilized in similar models~\cite{similar}.

We search for strongly-coupled gauge theories that can be employed to realize the above framework.
We consider QCD-like gauge theories with fermions as candidate theories,
 and assume that strong dynamics of the theory triggers confinement of the fermions and dynamical breaking of the flavor symmetry, analogously to QCD.
The candidates are classified into three categories, where
\\

(i) the fundamental representaion of the gauge group is complex, and there are $N_f$ left-handed fermions in the fundamental representation and $N_f$ in its conjugate representation.
The mesons resulting from confinement (including those which become Nambu-Goldstone (NG) bosons and others) are in $N_f \times N_f$ representation of $SU(N_f)_1 \times SU(N_f)_2$ flavor symmetry,
 and along dynamical symmetry breaking, it breaks as $SU(N_f)_1 \times SU(N_f)_2 \to SU(N_f)$.

(ii) the fundamental representaion of the gauge group is real, and there are $2N_f$ Weyl fermions in the fundamental representation.
The mesons are in the rank-2 symmetric representation of $SU(2N_f)$ flavor symmetry,
 and along dynamical symmetry breaking, it breaks as $SU(2N_f) \to SO(2N_f)$.

(iii) the fundamental representaion of the gauge group is pseudo-real, and there are $2N_f$ Weyl fermions in the fundamental representation;
The mesons are in the rank-2 antisymmetric representation of $SU(2N_f)$ flavor symmetry,
 and along dynamical symmetry breaking, the symmetry breaks as $SU(2N_f) \to USp(2N_f)$.
\\

\noindent
For the model building, we impose two requirements below on the candidate gauge theories:
\\

(a) It should be possible to embed the $SU(2)_W\times U(1)_Y$ electroweak symmetry into the flavor symmetry in such a way that
 it is not broken along dynamical symmetry breaking and that there exists a meson in (2, $1/2$) representation.

(b) All the NG bosons should be charged under the $SU(2)_W\times U(1)_Y$ electroweak gauge group, so that they gain mass
 through electroweak interactions.
\\

\noindent
(a) is mandatory to realize the mixing of a meson with an elementary scalar that yields the SM Higgs field 
 (we do not consider cases where a bosonic baryon, instead of a meson, mixes to give the SM Higgs field).
(b) is necessary to construct an experimentally viable model; NG bosons neutral under the $SU(2)_W\times U(1)_Y$ group
 are massless, as current mass is absent due to classical scale invariance, or even become tachyonic through Yukawa interaction with the elementary scalar $H$.
Moreover, these NG bosons have Wess-Zumino-Witten term~\cite{wzw} with electroweak gauge bosons and could be accessed in collider experiments.
The presence of such bosons would impose a strong restriction on the dynamical scale and hence on the mixing angle of the elementary and composite scalars, rendering the MPP conditions unachievable.

As a matter of fact, (a) and (b) cannot be met in all of the type-(i), (ii) and (iii) theories.
Nevertheless, we find that introduction of a new weakly-coupled gauge symmetry into type-(iii) theory with $N_f=2$ can reach the goal.
The model we consider is based on a strongly-coupled gauge theory with 4 Weyl fermions in its pseudo-real fundamental representation,
 where the $SU(2)_W\times U(1)_Y$ symmetry is straightforwardly embedded into the $SU(4)$ flavor symmetry to have a meson in (2, $1/2$) representation.
Along dynamical symmetry breaking, it breaks as $SU(4) \to USp(4)$, with $USp(4) \supset SU(2)_W\times U(1)_Y$ and hence the electroweak symmetry maintained.
There appear 5 NG bosons, 4 being charged under $SU(2)_W\times U(1)_Y$ and 1 being neutral.
The key is to gauge part of the flavor symmetry corresponding to this neutral NG boson, which we call $U(1)_X$ gauge symmetry, 
 so that the neutral NG boson becomes the longitudinal component of the $U(1)_X$ gauge boson along dynamical symmetry breaking, like in the technicolor model~\cite{tc}, and does not appear as a light physical field.
To cancel $U(1)_X - SU(2)_W - SU(2)_W$, $U(1)_X - U(1)_Y - U(1)_Y$, $U(1)_X - U(1)_X - U(1)_Y$ and $U(1)_X - U(1)_X - U(1)_X$ chiral anomalies,
 we assign $U(1)_X$ charges not only to the fermions in the strongly-coupled gauge theory but also to SM fermions,
 which is successful only in the current model with $SU(4) \to USp(4)$ breaking.
For concreteness, in this paper, we choose the minimal gauge group for the type-(iii) strongly-coupled gauge theory, that is, $SU(2)$.

This paper is organized as follows:
After the introduction, we describe in detail the field content and gauge symmetry of the model,
 and study $SU(2)$ gauge dynamics to show that the mixing of elementary and composite scalars can yield the SM Higgs field.
Next, by a numerical analysis on the RG equations for the coupling constants, we demonstrate that the MPP conditions can be satisfied
 for some values of the top quark pole mass and the mixing angle of the elementary and composite scalars.
We further derive the spectrum of light new particles in the model and discuss its phenomenological implications,
 and then conclude the paper.
\\

The gauge symmetry of the model is $SU(3)_C \times SU(2)_W \times U(1)_Y \times SU(2)_T \times U(1)_X$, where
 $SU(3)_C$, $SU(2)_W$ and $U(1)_Y$ are the SM color, weak and hypercharge gauge groups, respectively, while $SU(2)_T$ and $U(1)_X$ are newly gauge groups.
The $SU(2)_T$ gauge theory becomes strongly-coupled at infrared scales and plays an essential role in the model,
 while the $U(1)_X$ gauge group is introduced to avoid having an extremely light NG boson after dynamical symmetry breaking in the $SU(2)_T$ gauge theory.
The model contains the SM fermions plus three flavors of SM-gauge-singlet neutrinos,
 and new fermions charged under the electroweak and new gauge groups $SU(2)_T \times U(1)_X \times SU(2)_W \times U(1)_Y$.
Further contained is an elementary scalar field with the same electroweak charge as the SM Higgs field and without $SU(2)_T$ or $U(1)_X$ charge.
The field content is shown in Table~\ref{content}.
\begin{table}[h]
\begin{center}
\begin{tabular}{|c||c|c|c|c|c|c|c|} \hline
    & Lorentz $SO(1,3)$ & $SU(3)_C$ & $SU(2)_W$ & $U(1)_Y$ & $SU(2)_T$ & $U(1)_X$ & flavor\\ \hline
$q$ & \bf{(1,2)}   & \bf{3}& \bf{2}          & $+1/6$ & \bf{1}& $x$ & 3 \\ 
$u^c$ & \bf{(1,2)} & \bf{$\bar{3}$} & \bf{1} & $-2/3$ & \bf{1}& $-x$ & 3 \\ 
$d^c$ & \bf{(1,2)} & \bf{$\bar{3}$} & \bf{1} & $+1/3$ & \bf{1}& $-x$ & 3 \\ 
$\ell$ & \bf{(1,2)} & \bf{1} & \bf{2}        & $-1/2$ & \bf{1}& $-3x-2/3$ & 3 \\ 
$e^c$  & \bf{(1,2)} & \bf{1} & \bf{1}        & $+1$   & \bf{1}& $3x+2/3$ & 3 \\ 
$n^c$  & \bf{(1,2)} & \bf{1} & \bf{1}        & $0$    & \bf{1}& $3x+2/3$ & 3 \\ \hline
$\chi=(\chi_1,\chi_2)$ & \bf{(1,2)} & \bf{1}   & \bf{2} & $0$   & \bf{2}& $+1$ & - \\ 
$\psi_1$ & \bf{(1,2)} & \bf{1} & \bf{1}      & $+1/2$ & \bf{2}& $-1$ & - \\ 
$\psi_2$ & \bf{(1,2)} & \bf{1} & \bf{1}      & $-1/2$ & \bf{2}& $-1$ & - \\ \hline
$H$      & \bf{1}     & \bf{1} & \bf{2}      & $+1/2$ & \bf{1}& $0$ & - \\ \hline
\end{tabular}
\end{center}
\caption{Field content of the model. Also shown are the Lorentz transformation properties and charge assignments 
 in the $SU(3)_C \times SU(2)_W \times U(1)_Y \times SU(2)_T \times U(1)_X$ gauge group.
Here, $x$ can be an arbitrary number.
}
\label{content}
\end{table}
Note in particular that chiral anomaly involving the electroweak gauge symmetry and the $U(1)_X$ is absent.
The $SU(2)_T$ gauge theory involves 4 Weyl fermions, $\chi=(\chi_1,\chi_2)$, $\psi_1$ and $\psi_2$.
For notational convenience, we write them interchangeably as
\begin{align}
\left(
\begin{array}{c}
\chi_1  \\
\chi_2  \\
\psi_1  \\
\psi_2
\end{array}
\right)
&=\left(
\begin{array}{c}
\chi  \\
\psi
\end{array}
\right)
=\Psi.
\end{align}
Classical scale invariance forbids mass term for the elementary scalar field $H$.
There is a Yukawa-type coupling among $H$, $\chi$, $\psi_1$ and $\psi_2$, given by
\begin{align}
-{\cal L}_{H\chi\psi} &= -y_1 \, H^\dagger \, \psi_1^T \epsilon_s\epsilon_t \chi - y_2 \, (H^T\epsilon_w^T) \, \psi_2 \epsilon_s\epsilon_t \chi
- y_1^* \, \chi^\dagger \epsilon_s\epsilon_t \psi_1^* \, H - y_2^* \, \chi^\dagger \epsilon_s\epsilon_t \psi_2^* \, (\epsilon_wH^*)
\nonumber \\
&= {\rm tr}\left[ \, 
\left(
\begin{array}{cc}
y_1 & 0 \\
0 & y_2
\end{array}
\right)
\left(
\begin{array}{c}
H^\dagger \\
H^T \epsilon_w^T
\end{array}
\right)
\chi \epsilon_s \epsilon_t \psi^T
+
\psi^* \epsilon_s \epsilon_t \chi^\dagger
\left(
\begin{array}{cc}
 H & \epsilon_w H^*
\end{array}
\right)
\left(
\begin{array}{cc}
y_1^* & 0 \\
0 & y_2^*
\end{array}
\right)
 \, \right],
\label{yukawa}
\end{align}
 where $\epsilon_s$, $\epsilon_t$ and $\epsilon_w$ denote the antisymmetric tensors acting on spinor indices, $SU(2)_T$ gauge indices, and $SU(2)_W$ gauge indices, respectively,
 and it is granted that $\chi \epsilon_s \epsilon_t \psi^T$ represents a 2$\times$2 matrix composed of bilinears of fermions $\chi_1,\chi_2,\psi_1,\psi_2$
 and $\psi^* \epsilon_s \epsilon_t \chi^\dagger$ represents its hermitian conjugate.
Additionally, we have a quartic coupling for the elementary scalar $H$ and Yukawa couplings for $H$, SM fermions and SM-gauge-singlet neutrinos, expressed as
\begin{align}
-{\cal L}_{{\rm quartic+Yukawa}} &=
\lambda \, ( H^\dagger H )^4 + Y_u \, \bar{q}u \, (\epsilon_w H^*) + Y_d \, \bar{q}d \, H + Y_n \, \bar{\ell}n \, (\epsilon_w H^*) + Y_e \, \bar{\ell}e \, H + {\rm h.c.},
\label{quartic+yukawa}
\end{align}
 where flavor indices are omitted.
The above term induces the SM Higgs quartic coupling, the SM Yukawa couplings and the Dirac Yukawa coupling for neutrino mass
 after $H$ mixes with mesons in the $SU(2)_T$ gauge theory.

The $SU(2)_T$ gauge theory possesses $SU(4)$ global symmetry at quantum level.
We label its 15 generators in the basis $(\chi_1, \, \chi_2, \, \psi_1, \, \psi_2)$ as
\begin{align}
T^a &= \frac{1}{2}\left(
\begin{array}{cc}
\sigma^a & O \\
O & O
\end{array}
\right), \
T^{a+3} = \frac{1}{2}\left(
\begin{array}{cc}
O & O \\
O & \sigma^a
\end{array}
\right), \ 
T^{a+6} = \frac{1}{2\sqrt{2}}\left(
\begin{array}{cc}
O & \sigma^a \\
\sigma^a & O
\end{array}
\right), \ 
T^{a+9} = \frac{1}{2\sqrt{2}}\left(
\begin{array}{cc}
O & -i\sigma^a \\
i\sigma^a & O
\end{array}
\right), 
\nonumber \\
T^{13} &= \frac{1}{2\sqrt{2}}\left(
\begin{array}{cc}
O & I \\
I & O
\end{array}
\right), \ 
T^{14} = \frac{1}{2\sqrt{2}}\left(
\begin{array}{cc}
O & -i I \\
i I & O
\end{array}
\right), \ 
T^{15} = \frac{1}{2\sqrt{2}}\left(
\begin{array}{cc}
I & O \\
O & -I
\end{array}
\right) \ \ \ \ \ (a=1,2,3),
\end{align}
 where $\sigma^a$'s are Pauli matrices, and $I$ and $O$ denote $2\times2$ unit and zero matrices, respectively.
$T^1,T^2,T^3$ are the generators for the weak gauge group $SU(2)_W$
 and $T^6$ is that for the hypercharge gauge group $U(1)_Y$.
$T^{15}$ is the generator for the new gauge group $U(1)_X$.

We assume that the $SU(2)_T$ gauge theory becomes strongly-coupled at infrared scales and triggers confinement and dynamical symmetry breaking as in QCD.
It is also assumed that the $U(1)_X$ gauge coupling is smaller than the weak and hypercharge gauge couplings.
Then, given the most attractive channel hypothesis~\cite{mac}, the dynamical symmetry breaking occurs in the pattern that preserves the $SU(2)_W \times U(1)_Y$ electroweak gauge symmetry but breaks the $U(1)_X$ gauge symmetry,
 which is given by
\begin{align}
\langle 0 \vert \Psi^T\epsilon_s \epsilon_t \, E \, \Psi \vert 0 \rangle &\neq 0, \ \ \ \ \ E \equiv \left(
\begin{array}{cccc}
0 & 1& 0& 0 \\
-1 & 0& 0& 0 \\
0 & 0& 0& 1 \\
0 & 0& -1& 0
\end{array}
\right).
\label{dsb}
\end{align}
In the dynamical symmetry breaking, the $SU(4)$ global symmetry is spontaneously broken into $USp(4)$ symmetry,
 along which the generators $T^{a+9}$ $(a=1,2,3)$, $T^{13}$ and $T^{15}$ are broken.
Since $T^{15}$ is the generator for the $U(1)_X$ gauge group, the NG boson associated with its breaking
 becomes the longitudinal component of the $U(1)_X$ gauge boson by the Higgs mechanism, as in the technicolor model~\cite{tc}.
The NG bosons associated with $T^{a+9}$ $(a=1,2,3)$ and $T^{13}$, denoted by $\Pi^b$ $(b=10,11,12,13)$, appear as physical fields
 and couple to the currents in the following way:
\begin{align}
\langle 0 \vert \, \Psi^\dagger \sigma_\mu T^{b} \Psi \, \vert \Pi^{c} \rangle &= i f_\Pi \, p_\mu \, \delta^{bc} \ \ \ \ \ (b,c=10,11,12,13),
\label{ngcoupling}
\end{align}
 where $f_\Pi$ is the NG boson decay constant, approximated to be common for all NG bosons.
Since electroweak gauge interactions explicitly violate $T^{b}$ $(b=10,11,12,13)$ part of the $SU(4)$ global symmetry,
 the NG bosons $\Pi^b$ are pseudo-NG (pNG) bosons with mass, whose origin is identical with the mass difference between the charged and neutral pions in QCD.
Their mass, $M_{\Pi^b}$, is computed with Dashen's formula~\cite{dashen} as
\begin{align}
M_{\Pi^b}^2 &= \frac{1}{f_\Pi^2}\langle 0 \vert [ Q^b, [Q^b, {\cal H}_{{\rm break}}]] \vert 0 \rangle,
\label{dashen}
\end{align}
 where $Q^b$ is the conserved charge for the current of generator $T^b$, and ${\cal H}_{{\rm break}}$ is the effective Hamiltonian that explicitly breaks the $SU(4)$ global symmetry.
In the leading order of the electroweak gauge couplings, and when the coupling constants $y_1,y_2$ and the electroweak symmetry breaking are ignored,
 ${\cal H}_{{\rm break}}$ reads
\begin{align}
&{\cal H}_{{\rm break}} = -\frac{i}{2} \, \frac{g_W^2}{4} \int{\rm d}^4x \, D_{\mu\nu}(x) \, \sum_{c=1}^3 \, \chi^\dagger(x)\sigma^{\mu}\sigma^c\chi(x) \ \chi^\dagger(0)\sigma^{\nu}\sigma^c\chi(0) 
\nonumber \\
&- \frac{i}{2} \frac{g_Y^2}{4} \int{\rm d}^4x D_{\mu\nu}(x)
\{ \psi_1^\dagger(x)\sigma^{\mu}\psi_1(x) - \psi_2^\dagger(x)\sigma^{\mu}\psi_2(x) \} 
\{ \psi_1^\dagger(0)\sigma^{\mu}\psi_1(0) - \psi_2^\dagger(0)\sigma^{\mu}\psi_2(0) \}
\label{effh}
\end{align}
 where $D_{\mu\nu}$ is the propagator for a free massless gauge field,
 and $g_W$ and $g_Y$ are the weak and hypercharge gauge couplings, respectively.
Substituting Eq.~(\ref{effh}) into Eq.~(\ref{dashen}), one obtains
\begin{align}
M_{\Pi^b}^2 &= \frac{i}{2} \, \frac{1}{f_\Pi^2}\int{\rm d}^4x \, 
\left( \frac{g_W^2}{4}+\frac{g_Y^2}{4} \right) D_{\mu\nu}(x)
\langle 0 \vert T\left\{ \chi_1^\dagger(x)\sigma^\mu \chi_1(x) \, \chi_2^T(0)\epsilon_s^T \bar{\sigma}^\nu \epsilon_s\chi_2^*(0) \right\} \vert 0 \rangle,
\nonumber \\
&= \frac{i}{2} \, \frac{1}{f_\Pi^2}\int{\rm d}^4x \, 
\left( \frac{g_W^2}{4}+\frac{g_Y^2}{4} \right) D_{\mu\nu}(x)
\langle 0 \vert T\left\{ \psi_1^\dagger(x)\sigma^\mu \psi_1(x) \, \psi_2^T(0)\epsilon_s^T \bar{\sigma}^\nu \epsilon_s\psi_2^*(0) \right\} \vert 0 \rangle \equiv M_\Pi^2,
\label{pngmass}
\end{align}
 where we exploit the fact that the $SU(2)_T$ gauge dynamics does not distinguish $\chi_1,\chi_2,\psi_1,\psi_2$ and that correlators of two $\chi_i$ fields and two $\psi^\dagger_j$ fields vanish,
 and, as $M_{\Pi^b}^2$'s are common, they are rewritten as $M_{\Pi}^2$.
We stress that the correlator in Eq.~(\ref{pngmass}) is proportional to the square of dynamical symmetry breaking VEV $\langle 0 \vert \chi_1^T \epsilon_s \epsilon_t \chi_2 \vert 0 \rangle = \langle 0 \vert \psi_1^T \epsilon_s \epsilon_t \psi_2 \vert 0 \rangle \neq 0$.
Eq.~(\ref{pngmass}) is compared to an analogous formula for the mass difference between charged and neutral pions,
\begin{align}
m_{\pi^{\pm}}^2 - m_{\pi^0}^2 &= \frac{i}{2} \, \frac{1}{f_\pi^2}\int{\rm d}^4x \, 4e^2 D^\gamma_{\mu\nu}(x)
\langle 0 \vert T\left\{ q_L^\dagger(x)\sigma^\mu q_L(x) \, q_R(0)^\dagger \bar{\sigma}^\nu q_R(0) \right\} \vert 0 \rangle,
\label{massdiff}
\end{align}
 where $q_L$, $q_R$ respectively denote left-handed and right-handed quarks, $e$ denotes electromagnetic coupling and $f_\pi$ denotes the pion decay constant.
The correlator in Eq.~(\ref{massdiff}) is proportional to the square of chiral symmetry breaking VEV $\langle0\vert q_L^\dagger q_R \vert 0\rangle \neq 0$.
Defining the ratio of the dynamical scales of the $SU(2)_T$ gauge theory and QCD as $r \equiv \Lambda_T/\Lambda_{QCD}$ and assuming it to be the same as the dynamical symmetry breaking VEV ratio,
 we obtain the following expression for the pNG boson mass:
\begin{align}
M_{\Pi}^2 &= r^2 \frac{g_W^2+g_Y^2}{16e^2} \frac{f_\pi^2}{(f_\pi^{{\rm chiral}})^2} (m_{\pi^{\pm}}^2 - m_{\pi^0}^2)
\label{pngmass2}
\end{align}
 where $f_\pi^{{\rm chiral}}$ denotes the pion decay constant in chiral-limit QCD.
From experimental values~\cite{pdg} and a lattice calculation of $f_\pi^{{\rm chiral}}$~\cite{durr}, we find
\begin{align}
M_{\Pi}^2 &= r^2 \, (0.0231 \, {\rm GeV})^2.
\label{pngmass3}
\end{align}

We derive asymptotic expressions for $\Pi^{b}$ $(b=10,11,12,13)$ in terms of fundamental fermions $\Psi$.
For this purpose, we remind that the pNG bosons should transform, under the unbroken generators $T^{a}$, $T^{a+3}$, $T^{a+6}$ $(a=1,2,3)$ and $T^{14}$, as adjoint representations corresponding to $T^{b}$.
We further note that under charge-conjugation-parity transformation, ${\cal CP}$, followed by a global $SU(4)$ transformation, ${\cal J}_2=e^{i(\pi/2)(T^2+T^5)}$,
 each side of Eq.~(\ref{ngcoupling}) transforms as
\begin{align}
&\langle 0 \vert \, ({\cal J}_2 \cdot {\cal CP}) \, \Psi^\dagger \sigma_\mu T^{b} \Psi \, ({\cal J}_2 \cdot {\cal CP})^{-1} \, ({\cal J}_2 \cdot {\cal CP}) \, \vert \Pi^{c} \rangle = i f_\Pi \, p_\mu \, \delta^{bc}
\nonumber \\
&\Rightarrow 
\langle 0 \vert \, (-1) \Psi^\dagger \sigma^\mu T^{b} \Psi \, ({\cal J}_2 \cdot {\cal CP} \vert \Pi^{c} \rangle) = i f_\Pi \, p^\mu \, \delta^{bc}, \ \ \ \ \ (b,c=10,11,12,13),
\label{ngcouplingtrans}
\end{align}
 which implies that the pNG bosons should transform as $\Pi^b \rightarrow -\Pi^b$ under the ${\cal J}_2 \cdot {\cal CP}$ transformation.
These two requirements uniquely fix the asymptotic expressions for the pNG bosons to be
\footnote{
To see that Eq.~(\ref{asym1}) transforms as adjoint representations corresponding to $T^{b}$,
 use an identity for the unbroken generators $T^{\hat{a}}$ $(\hat{a}=1,2,...,9,14)$, $(T^{\hat{a}})^T E = -E T^{\hat{a}}$.
To verify that Eq.~(\ref{asym1}) is odd under the ${\cal J}_2 \cdot {\cal CP}$ transformation, note the fact that $\Psi$ has $\pm i$ eigenvalue in the ${\cal CP}$ transformation.
}
\begin{align}
\Pi^{b} &\propto \frac{1}{2} \left( \, \Psi^T\epsilon_s \epsilon_t \, E T^{b} \,
\Psi - \Psi^\dagger\epsilon_s \epsilon_t \, T^{b} E \, \Psi^* \,
\right) \ \ \ \ \ (b=10,11,12,13).
\label{asym1}
\end{align}
The remaining components of $\Psi^T\epsilon_s \epsilon_t \, E T^{b} \, \Psi$, which are even under the ${\cal J}_2 \cdot {\cal CP}$ transformation,
 correspond to asymptotic expressions for another set of mesons, $\Theta^b$ $(b=10,11,12,13)$, that transforms as $\Theta^b \to \Theta^b$ under the ${\cal J}_2 \cdot {\cal CP}$ transformation.
Namely, we have
\begin{align}
\Theta^{b} &\propto \frac{1}{2i} \left( \, \Psi^T\epsilon_s \epsilon_t \, E T^{b} \, \Psi + \Psi^\dagger\epsilon_s \epsilon_t \, T^{b} E \, \Psi^* \,
\right) \ \ \ \ \ (b=10,11,12,13).
\end{align}
Since $\Theta^b$'s are not NG bosons, they gain mass at the dynamical scale of the $SU(2)_T$ gauge theory,
 which we approximate to be common and denote by $M_\Theta$.
Now that we have obtained asymptotic expressions for $\Pi^b$ and $\Theta^b$,
 we formulate their scalar decay constants, $G_\Pi$ and $F_\Theta$, which we approximate to be common for $b=10,11,12,13$, as
\begin{align}
\langle 0 \vert \, \frac{1}{2} \left( \, \Psi^T\epsilon_s \epsilon_t \, E T^{b} \,
\Psi - \Psi^\dagger\epsilon_s \epsilon_t \, T^{b} E \, \Psi^* \,
\right) \, \vert \Pi^{c} \rangle &= G_\Pi \, \delta^{bc},
\label{scalardecayconst1} \\
\langle 0 \vert \, \frac{1}{2i} \left( \, \Psi^T\epsilon_s \epsilon_t \, E T^{b} \,
\Psi + \Psi^\dagger\epsilon_s \epsilon_t \, T^{b} E \, \Psi^* \,
\right) \, \vert \Theta^{c} \rangle &= F_\Theta M_\Theta \, \delta^{bc}.
\label{scalardecayconst2}
\end{align}
It is insightful to define the following canonically-normalized $SU(2)_W$ doublet fields with hypercharge $Y=1/2$, $\Pi$ and $\Theta$:
\begin{align}
\Pi \equiv \frac{1}{\sqrt{2}}\left(
\begin{array}{c}
\Pi^{11}+i\Pi^{10}  \\
\Pi^{13}-i\Pi^{12} 
\end{array}
\right), \ \ \ \ \ 
\Theta &\equiv \frac{1}{\sqrt{2}}\left(
\begin{array}{c}
\Theta^{11}+i\Theta^{10}  \\
\Theta^{13}-i\Theta^{12} 
\end{array}
\right),
\end{align}
 with which the asymptotic expressions for the NG bosons and massive mesons take the following form:
\begin{align}
G_\Pi\left(
\begin{array}{cc}
\epsilon_w \Pi^* & \Pi
\end{array}
\right)
\left(
\begin{array}{cc}
0 & 1 \\
-1 & 0
\end{array}
\right)
+ i \, F_\Theta M_\Theta\left(
\begin{array}{cc}
\epsilon_w \Theta^* & \Theta
\end{array}
\right)
\left(
\begin{array}{cc}
0 & 1 \\
-1 & 0
\end{array}
\right)
&= \chi \epsilon_s \epsilon_t \psi^T.
\end{align}

The Yukawa-type coupling Eq.~(\ref{yukawa}) induces mixing terms among the elementary scalar field $H$, the NG boson $\Pi$ and the massive meson $\Theta$, given by
\begin{align}
-{\cal L}_{H\chi\psi} &= i \, {\rm tr}\left[ \, \left(
\begin{array}{cc}
y_1 & 0 \\
0 & y_2
\end{array}
\right)
\left(
\begin{array}{c}
 H^\dagger  \\
 H^T \epsilon_w^T
\end{array}
\right)
\left\{ \,
G_\Pi
\left(
\begin{array}{cc}
\epsilon_w \Pi^* & \Pi
\end{array}
\right)
+ i \, F_\Theta M_\Theta\left(
\begin{array}{cc}
\epsilon_w \Theta^* & \Theta
\end{array}
\right)
\, \right\} 
\left(
\begin{array}{cc}
0 & 1 \\
-1 & 0
\end{array}
\right) \right.
\nonumber \\
&-
\left.
\left(
\begin{array}{cc}
0 & -1 \\
1 & 0
\end{array}
\right)
\left\{ \,
G_\Pi\left(
\begin{array}{c}
\Pi^T \epsilon_w^T \\
\Pi^\dagger
\end{array}
\right)
- i \, F_\Theta M_\Theta\left(
\begin{array}{c}
\Theta^T \epsilon_w^T \\
\Theta^\dagger
\end{array}
\right)
\, \right\}
\left(
\begin{array}{cc}
H & \epsilon_w H^*
\end{array}
\right)
\left(
\begin{array}{cc}
y_1^* & 0 \\
0 & y_2^*
\end{array}
\right) \, 
\right]
\nonumber \\
&= (y_1-y_2^*)F_\Theta M_\Theta \, \Theta^\dagger H + (y_1^*-y_2)F_\Theta M_\Theta \, H^\dagger \Theta + i(y_1^*+y_2) G_\Pi \, \Pi^\dagger H - i(y_1+y_2^*) G_\Pi \, H^\dagger \Pi.
\end{align}
With the above mixing terms, the mass matrix for the elementary scalar $H$, the NG boson $\Pi$ and the massive meson $\Theta$ is derived to be
\begin{align}
-{\cal L} &\supset \left(
\begin{array}{ccc}
H^\dagger & \Theta^\dagger & \Pi^\dagger
\end{array}
\right)
\left(
\begin{array}{ccc}
0 & (y_1^*-y_2) F_\Theta M_\Theta & -i(y_1+y_2^*) G_\Pi \\
(y_1-y_2^*) F_\Theta M_\Theta & M_\Theta^2 & 0 \\
i(y_1^*+y_2) G_\Pi & 0 & M_\Pi^2
\end{array}
\right)
\left(
\begin{array}{c}
H  \\
\Theta  \\
\Pi
\end{array}
\right).
\label{massmatrix}
\end{align}
The mass matrix Eq.~(\ref{massmatrix}) can be further approximated:
Without fine-tuning between $y_1$ and $y_2$, we have $\vert y_1^*-y_2\vert F_\Theta M_\Theta \sim \vert y_1^* +y_2\vert G_\Pi$,
 because $F_\Theta M_\Theta$ and $G_\Pi$ have the same dynamical origin.
In contrast, we have $M_\Theta^2 \gg M_\Pi$,
 because the $\Theta$ meson mass is about the dynamical scale of the $SU(2)_T$ gauge theory,
 whereas the mass of $\Pi$ meson, which is a pseudo-NG boson, is suppressed by $g_W^2$ times a loop factor $1/(16\pi^2)$ compared to the dynamical scale, as is found in Eq.~(\ref{pngmass}).
Therefore, Eq.~(\ref{massmatrix}) can be approximated as
\begin{align}
-{\cal L} &\supset M_\Theta^2 \, \Theta^\dagger \Theta + \left(
\begin{array}{cc}
H^\dagger & \Pi^\dagger
\end{array}
\right)
\left(
\begin{array}{ccc}
0 & -i(y_1+y_2^*) G_\Pi \\
i(y_1^*+y_2) G_\Pi & M_\Pi^2
\end{array}
\right)
\left(
\begin{array}{c}
H  \\
\Pi
\end{array}
\right).
\label{massmatrixappr}
\end{align}
We further rotate the phase of $H$ to make $-i(y_1+y_2^*)$ real.
After diagonalization, the mass matrix becomes
\begin{align}
-{\cal L} &\supset M_\Theta^2 \, \Theta^\dagger \Theta - M_1^2 \, H_1^\dagger H_1 + M_2^2 \, H_2^\dagger H_2,
\label{massterms}
\end{align}
 where
\begin{align}
\left(
\begin{array}{c}
H  \\
\Pi
\end{array}
\right) &= 
\left(
\begin{array}{cc}
c_H & s_H \\
-s_H & c_H
\end{array}
\right)
\left(
\begin{array}{c}
H_1  \\
H_2
\end{array}
\right),
\ \ \ c_H = \sqrt{1-s_H^2} = \sqrt{\frac{M_\Pi^2+M_1^2}{M_\Pi^2+2M_1^2}},
\nonumber \\
-M_1^2+M_2^2&=M_\Pi^2, \ \ \ M_1^2M_2^2=\vert y_1^*+y_2\vert^2 G_\Pi^2, \ \ \ M_2^2>M_1^2>0.
\label{mixing}
\end{align}
Since $H_1$ has a negative mass squared term $-M_1^2$, it develops a non-zero VEV that breaks the electroweak symmetry.
We therefore identify $H_1$ with the SM Higgs field, which gives
\begin{align}
M_1^2 &= \frac{m_h^2}{2},
\label{m1mh}
\end{align}
 where $m_h\simeq125$~GeV is the SM Higgs particle mass.
The SM Higgs quartic coupling and the SM Yukawa couplings are induced from those of the elementary scalar $H$ in the fundamental Lagrangian Eq.~(\ref{quartic+yukawa}), as
\begin{align}
-{\cal L}_{{\rm quartic+Yukawa}} &\supset
\frac{\lambda}{c_H^4} \, ( H_1^\dagger H_1 )^4 + \frac{Y_u}{c_H} \, \bar{q}u \, (\epsilon_w H_1^*) + \frac{Y_d}{c_H} \, \bar{q}d \, H_1 + \frac{Y_n}{c_H} \, \bar{\ell}n \, (\epsilon_w H_1^*) + \frac{Y_e}{c_H} \, \bar{\ell}e \, H_1 + {\rm h.c.},
\label{inducedcouplings}
\end{align}
$\Pi$ possesses a quartic coupling that originates from the explicit breaking of the $SU(4)$ global symmetry by the electroweak gauge interaction.
However, it is roughly a loop factor $1/(16\pi^2)$ times $g_W^4$ and hence has a negligible contribution to the SM Higgs quartic coupling.
\\

We demonstrate that the quartic coupling for the elementary scalar $H$ satisfies
\begin{align}
\lambda(\mu \simeq M_P) &= 0, \ \ \ \ \ \beta_\lambda(\mu \simeq M_P) \equiv \frac{{\rm d}}{{\rm d}\log \mu} \lambda(\mu) \vert_{\mu\simeq M_P} = 0,
\end{align}
 ($\mu$ is a renormalization scale) for some values of the top quark pole mass and the mixing angle of $H$ and $\Pi$,
 realizing the MPP.
For this purpose, we numerically solve the RG equations for the $H$ quartic coupling $\lambda$
 and relevant coupling constants including the top quark Yukawa coupling $(Y_u)_{33}$ in Eq.~(\ref{quartic+yukawa}) and the SM gauge coupling constants.
The RG equations are obtained by adding contributions of new particles to the SM two-loop RG equations in Ref.~\cite{rge}.
We make the approximation that the particle content changes from (SM particles+pNG boson) to (SM particles+fermionic particles made of $\chi,\psi_1,\psi_2$ fields) 
 at some matching scale, $M_{{\rm match}}$, and ignore loop-level threshold corrections.
Hence, for renormalization scales $\mu < M_{{\rm match}}$, the pNG boson $\Pi$ contributes to the evolution of the weak and hypercharge gauge couplings.
At the scale $\mu=M_{{\rm match}}$, the SM Higgs quartic coupling, $\lambda^{SM}$, and top quark Yukawa coupling, $y_t^{SM}$, are
 matched to the $H$ quartic coupling $\lambda$ and the Yukawa coupling $(Y_u)_{33}$ as
\begin{align}
\frac{1}{c_H^4}\lambda(\mu=M_{{\rm match}}) &= \lambda^{{\rm SM}}(\mu=M_{{\rm match}}), \ \ \ \ \ \frac{1}{c_H}(Y_u)_{33}(\mu=M_{{\rm match}})=y_t^{{\rm SM}}(\mu=M_{{\rm match}}).
\end{align}
For scales $\mu > M_{{\rm match}}$, fermionic particles made from $\Psi$ affect the evolution of the weak and hypercharge gauge couplings.
A reasonable choice for the matching scale $M_{{\rm match}}$ is the $\Theta$ meson mass,
 because it corresponds to the confinement scale below which composite fields $\Theta$ as well as $\Pi$ appear.
In the analysis, therefore, we vary $M_{{\rm match}}$ about the $\Theta$ mass $M_\Theta$ as 
\begin{align}
M_\Theta/2 &\leq M_{{\rm match}} \leq 2 M_\Theta
\end{align}
 to examine the dependence on the matching scale.
We relate $M_\Theta$ to the pNG boson mass $M_\Pi$ and hence to $c_H$ based on analogy with QCD:
We argue that the $\Theta$ meson, being a massive two-fermion confining state, is most analogous to $K_0^*(1430)$ scalar meson in QCD~\cite{scalarmesons}.
Then $M_\Theta$ can be expressed in terms of the dynamical scale ratio $r=\Lambda_T/\Lambda_{QCD}$ and the $K_0^*(1430)$ mass as
 $M_\Theta = r \, m_{K_0^*(1430)}$, where $m_{K_0^*(1430)}=1.425$~GeV~\cite{pdg}.
Since $r$ is related to the pNG boson mass $M_\Pi$ through Eq.~(\ref{pngmass2}) and $M_\Pi$ is related to $c_H$ through Eqs.~(\ref{mixing}),~(\ref{m1mh}),
 $M_\Theta$ and $c_H$ are linked.
We fix SM parameters as $M_W=80.385$~GeV, $\alpha_s(M_Z)=0.1184$ and $m_h=125.09$~GeV,
 ignore contributions through Yukawa-type couplings $y_1,y_2$, and further assume the $U(1)_X$ gauge coupling to be negligibly small.

\begin{figure}[H]
  \begin{center}
    \includegraphics[width=80mm]{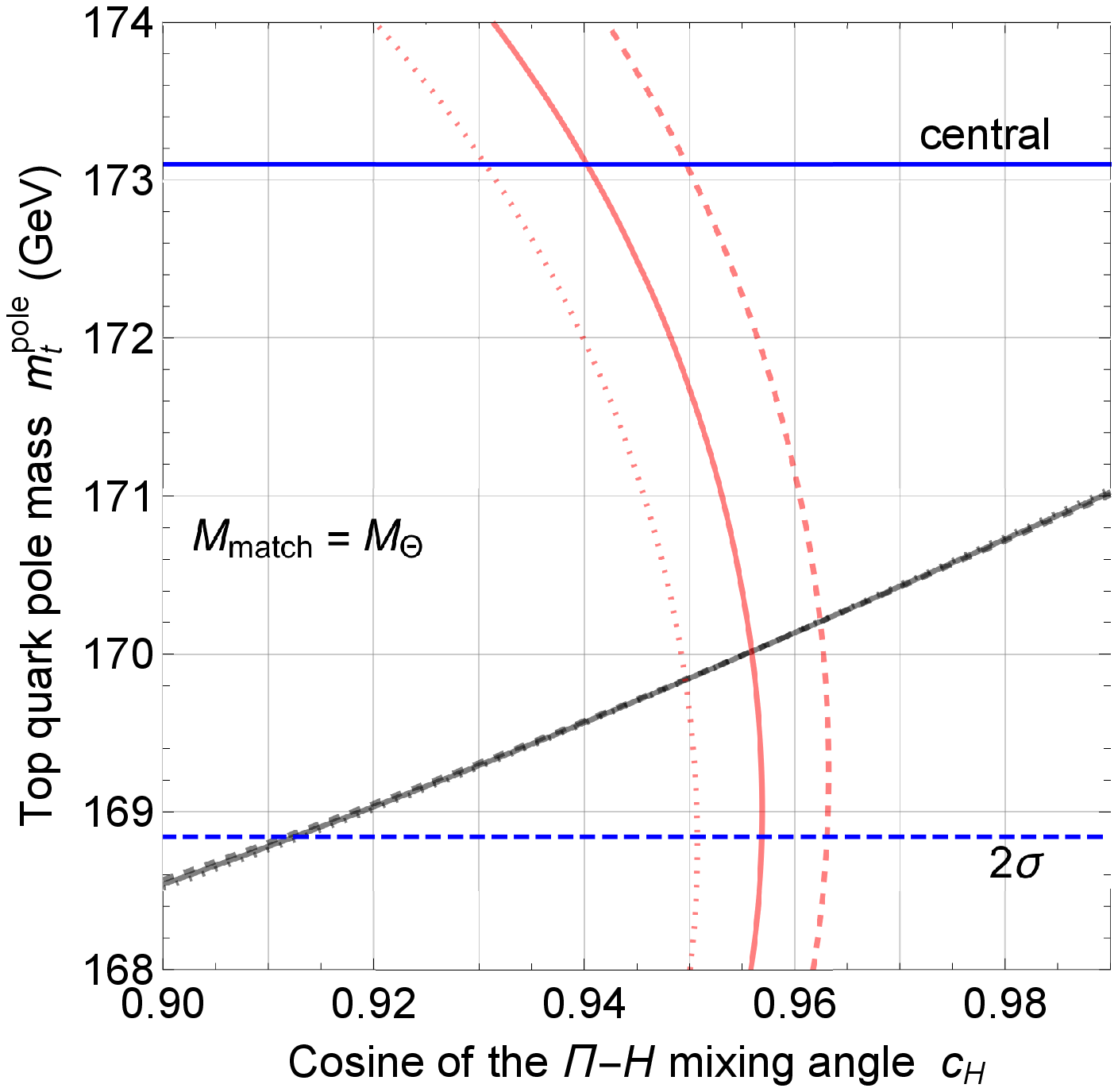}
    \\
    \includegraphics[width=80mm]{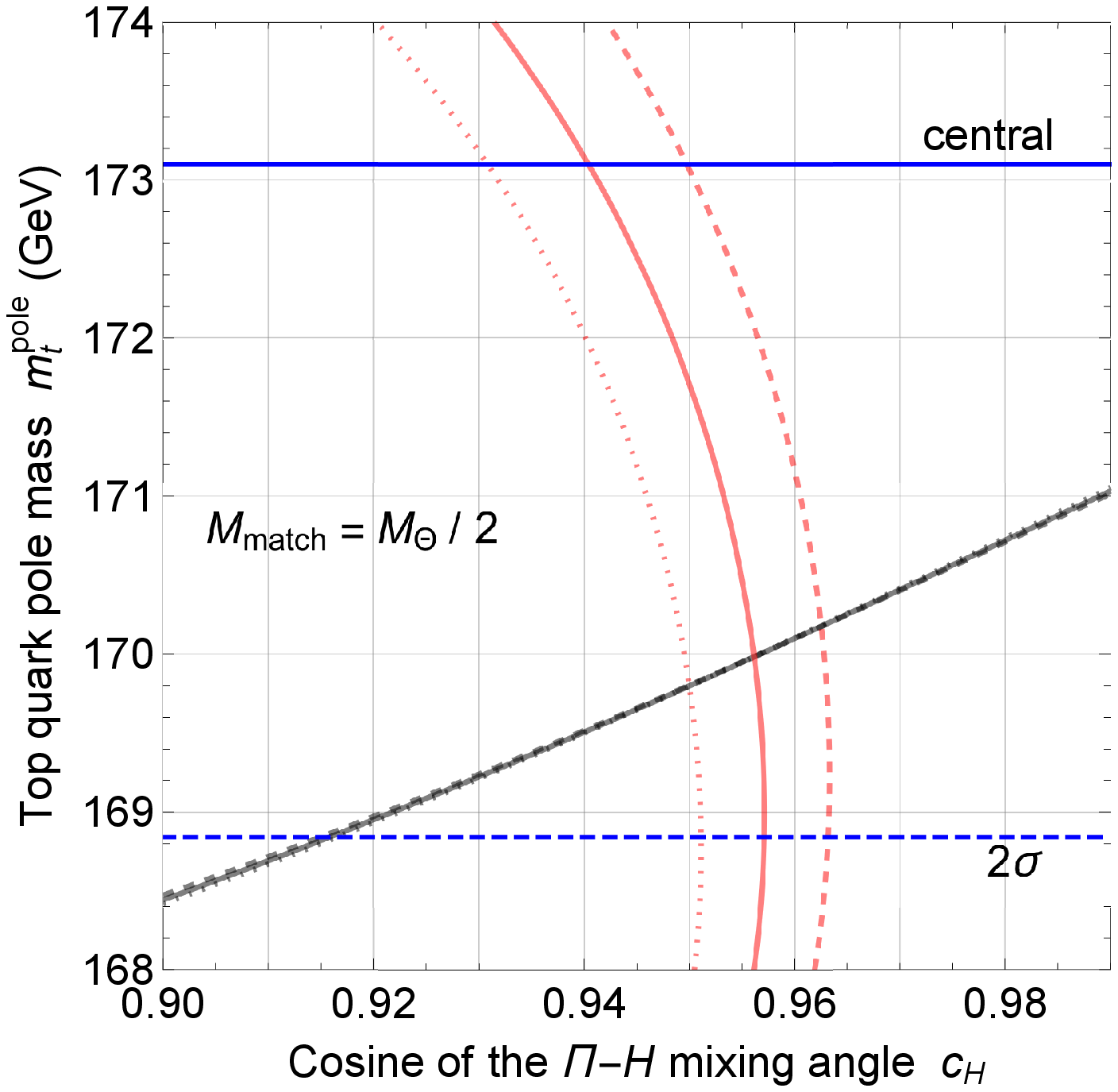}
    \includegraphics[width=80mm]{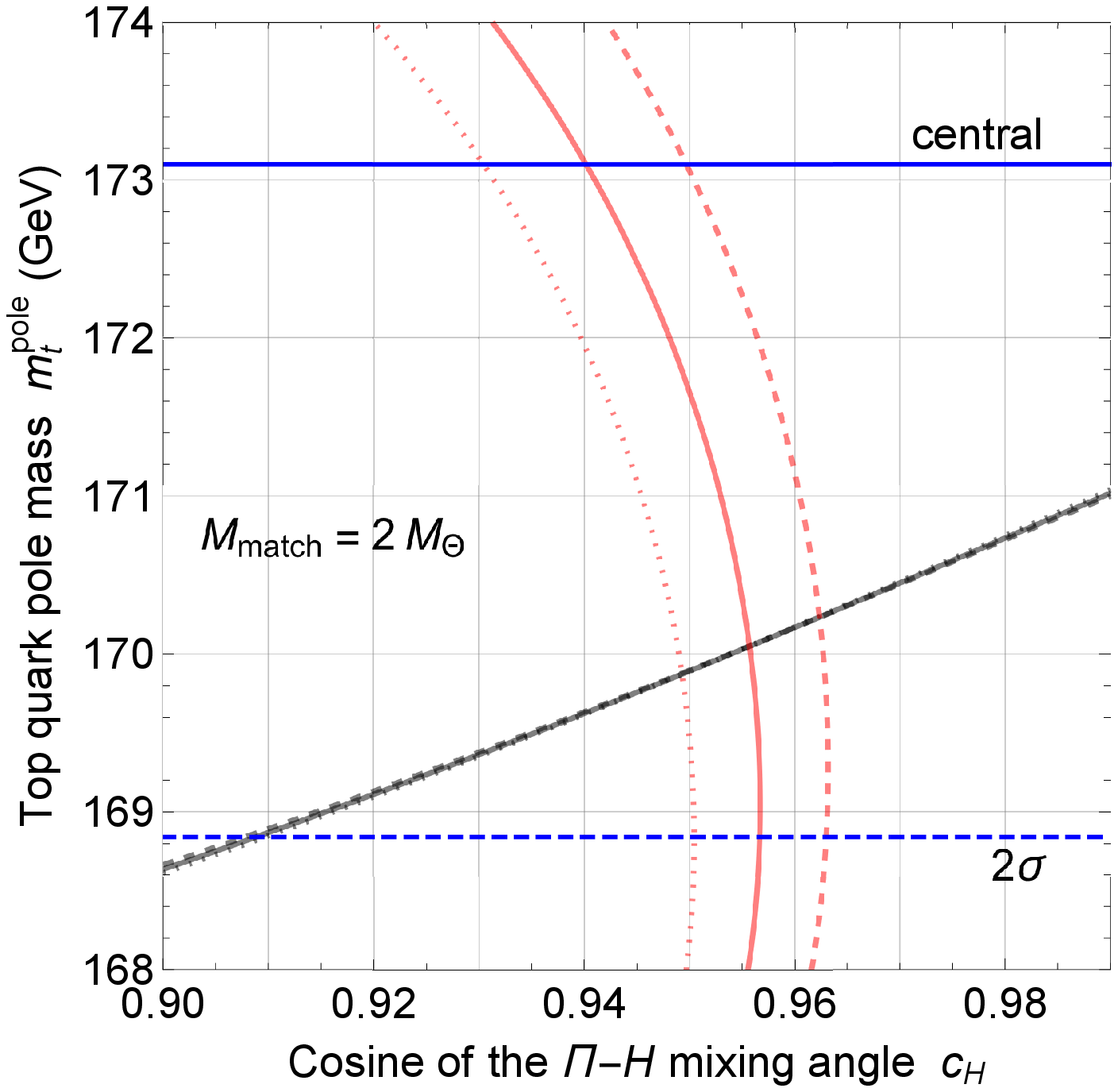}
    \caption{Contours for $\lambda(M_P/2)=0$, $\lambda(M_P)=0$ and $\lambda(2M_P)=0$, drawn by black-dashed, black-solid and black-dotted lines (almost degenerate in the plots),
    and contours for $\beta_\lambda(M_P/2)=0$, $\beta_\lambda(M_P)=0$ and $\beta_\lambda(2M_P)=0$, drawn by red-dashed, red-solid and red-dotted lines, respectively,
    on the plane spanned by the cosine of the mixing angle $c_H$ and the top quark pole mass $m_t^{{\rm pole}}$.
    The blue horizontal line 
    The left-bottom, up and right-bottom subplots correspond to different choices of the matching scale $M_{{\rm match}}$ in the RG equation
     with $M_{{\rm match}}=M_\Theta/2$, $M_{{\rm match}}=M_\Theta$ and $M_{{\rm match}}=2M_\Theta$, respectively.
    }
    \label{contours}
  \end{center}
\end{figure}

We present in Figure~\ref{contours} contours of $\lambda(M_P/2)=0$, $\lambda(M_P)=0$ and $\lambda(2M_P)=0$ by black-dashed, black-solid and black-dotted lines,
 and contours of $\beta_\lambda(M_P/2)=0$, $\beta_\lambda(M_P)=0$ and $\beta_\lambda(2M_P)=0$ by red-dashed, red-solid and red-dotted lines, respectively,
 on the plane spanned by the cosine of the mixing angle $c_H$ and the top quark pole mass $m_t^{{\rm pole}}$ 
 (the black-dashed, solid and dotted lines are nearly degenerate).
Each subplot corresponds to different matching scales, with $M_{{\rm match}}=M_\Theta/2$ for the left-bottom, $M_{{\rm match}}=M_\Theta$ for the up and $M_{{\rm match}}=2 M_\Theta$ for the right-bottom.
The blue-solid and dashed lines respectively indicate the central value and 2$\sigma$ lower bound for the top quark pole mass obtained from the pole mass direct measurement~\cite{atlastoppole}, which reports $m_t^{{\rm pole}}=173.1\pm2.1$~GeV.
The three subplots are similar, which assures us that the result is insensitive to the matching scale.
The intersections of the red and black lines indicate pairs of the mixing angle and top quark pole mass that realize the MPP.
We find that the cosine of the mixing angle $c_H$ and the top quark pole mass $m_t^{{\rm pole}}$ are given as
\begin{align}
{\rm for} \ M_\Theta/2 &< M_{{\rm match}} < 2 M_\Theta \ \ {\rm and} \ \ M_P/2 < ({\rm the \ scale \ where \ MPP \ is \ realized}) < 2 M_P,
\nonumber \\
0.949 &< c_H < 0.963, \ \ \ \ \ 169.8~{\rm GeV} < m_t^{{\rm pole}} < 170.3~{\rm GeV}.
\label{range}
\end{align}
The MPP conditions are satisfied with the Higgs particle mass of $m_h=125.09$~GeV and the top quark pole mass within 2$\sigma$ bound of its direct measurement~\cite{atlastoppole}.
We comment on experimental constraints from other top quark mass measurements~\cite{atlastop, cmstop}, which report $m_t=172.84\pm0.70$~GeV and $m_t=172.44\pm0.49$~GeV.
In these measurements, data are fit with a Monte Carlo simulation performed with an event generator
 which includes the top quark mass as a parameter, and some parameter value is regarded as the measured top quark mass.
Ref.~\cite{polevsmc} argues that the top quark pole mass can be smaller by 0.9~GeV compared to the top quark mass parameter in event generators.
Given this fact, the upper value of $m_t^{{\rm pole}}$ in our result Eq.~(\ref{range}) is adjacent to the 2$\sigma$ lower bounds of the measurements~\cite{atlastop, cmstop} owing to the 0.9~GeV separation.
\\

We study phenomenology of the model.
The mass of $H_2$ is related to $c_H$ as $M_2 = (c_H/\sqrt{1-c_H^2})(m_h/\sqrt{2})$ and hence takes values in the following range:
\begin{align}
266~{\rm GeV} &< M_2 < 316~{\rm GeV}.
\end{align}
The pNG boson mass is $M_\Pi^2=M_2^2-M_1^2=M_2^2-m_h^2/2$, which gives through Eq.~(\ref{pngmass3}) the dynamical scale ratio $r=\Lambda_T/\Lambda_{QCD}$.
Analogy with QCD yields the $\Theta$ meson mass $M_\Theta = r \, m_{K_0^*(1430)}$ and the NG boson decay costant $f_\Pi = \sqrt{2/3}r \, f_\pi^{{\rm chiral}}$, where $m_{K_0^*(1430)}$ and $f_\pi^{{\rm chiral}}$
 are the $K_0^*(1430)$ mass and the pion decay constant in chiral-limit QCD, respectively, and $\sqrt{2/3}$ accounts for difference in the gauge groups.
Using values in Refs.~\cite{pdg,durr}, we find
\begin{align}
1.55~{\rm TeV} &< M_\Theta < 1.87~{\rm TeV}, \ \ \ \ \ 770~{\rm GeV} < f_\Pi < 931~{\rm GeV}.
\end{align}
Note that $f_\Pi$ controls the mass of the $U(1)_X$ gauge boson, which originates from the dynamical symmetry breaking and can be calculated from Eq.~(\ref{ngcoupling}).
The spectrum of new particles below TeV scale thus comprises (i) isospin doublet scalar particles with hypercharge $+1/2$ made from $H_2$ field;
 (ii) the massive $U(1)_X$ gauge boson.
As the approximated mass matrix~Eq.~(\ref{massmatrixappr}) respects $CP$ symmetry,
 we label the charged, $CP$-even and $CP$-odd particles coming from $H_2$ field by $H_2^{\pm}$, $H_2^0$ and $A_2$, respectively.
$H_2$ field has Yukawa-type couplings with the SM fermions, which stem from those for the elementary scalar $H$ and are given in terms of the SM Yukawa couplings $y_u^{{\rm SM}},y_d^{{\rm SM}},y_e^{{\rm SM}}$
 and the neutrino Dirac Yukawa coupling $y_n$ as
\begin{align}
-{\cal L} &\supset \frac{s_H}{c_H}y_u^{{\rm SM}} \ \bar{q}u \, (\epsilon_w H_2^*) + \frac{s_H}{c_H}y_d^{{\rm SM}}  \ \bar{q}d \, H_2 + \frac{s_H}{c_H}y_n^{{\rm SM}}  \ \bar{\ell}n \, (\epsilon_w H_2^*) + \frac{s_H}{c_H}y_e^{{\rm SM}}  \ \bar{\ell}e \, H_2 + {\rm h.c.},
\label{inducedcouplings2}
\end{align}
 namely, the Yukawa couplings for $H_2$ are proportional to the SM ones with the suppression of $s_H/c_H (\sim 0.3)$.
On the other hand, since $H_2$ has no VEV, tree-level couplings among one $H_2$ and two $W/Z$ gauge bosons are absent.
Therefore, $H_2^\pm$, $H_2^0$ and $A_2$ particles mainly decay as $H_2^+ \to t \bar{b}$, $H_2^0/A_2 \to b \bar{b}$ and $H_2^0/A_2 \to \tau \bar{\tau}$ at tree level and $H_2^0/A_2 \to (gluon) (gluon)$ through a top quark loop,
 whereas they cannot decay into $W/Z$ bosons at tree level.
The branching fractions are found as
\begin{align}
&{\rm for \ }266~{\rm GeV} < M_2 < 316~{\rm GeV},
\nonumber \\
&Br(H_2^+ \to t \bar{b}) = 1.00,
\nonumber \\
&0.52 > Br(H_2^0/A_2 \to b \bar{b}) > 0.40, 
\ \ \ 0.066 > Br(H_2^0/A_2 \to \tau \bar{\tau}) > 0.057,
\nonumber \\
&0.41 < Br(H_2^0/A_2 \to (gluon) (gluon)) < 0.55,
\ \ \ 1.2\times10^{-3} < Br(H_2^0/A_2 \to \gamma \gamma) < 1.7\times10^{-3},
\end{align}
 which are based on calculations for a SM-like Higgs boson~\cite{wg}, with the contributions of tree-level $W/Z$ couplings removed.
Although tiny, the branching fraction for $H_2^0/A_2 \to \gamma \bar{\gamma}$ through a top quark loop is also shown because of its experimental importance.

In hadron collider experiments, promising channels to search for signals of $H_2^\pm$, $H_2^0$ and $A_2$ particles
 are the production of a single $H_2^0$ or $A_2$ particle through gluon fusion followed by the decay into $\tau \bar{\tau}$ or $\gamma \gamma$,
 the Drell-Yan production of a $H_2^\pm$ pair followed by the decay into $t\bar{b}\bar{t}b$,
 that of $H_2^0$ and $H_2^{\pm}$ pair or $A_2$ and $H_2^{\pm}$ pair followed by the decay into $\bar{b}b\bar{t}b/\bar{b}b\bar{b}t$ or $\bar{\tau}\tau\bar{t}b/\bar{\tau}\tau\bar{b}t$,
 and that of $H_2^0$ and $A_2$ pair followed by the decay into $\bar{b}b\bar{\tau}\tau$,
 since signals containing only bottom-quark jets in the final state are overwhelmed by QCD multijet background.
In 13~TeV proton-proton collisions, the cross section times branching fraction for each channel is calculated as
\begin{align}
&{\rm when \ }(c_H, \, M_2){\rm \ varies \ as \ }(0.949, \ 266~{\rm GeV}) < (c_H, \, M_2) < (0.963, \ 316~{\rm GeV}),
\nonumber \\
&68~{\rm fb} > \left( \, \sigma_{{\rm 13 \, TeV}}\left(p p \to H_2^0\right)+\sigma_{{\rm 13 \, TeV}}\left(p p \to A_2\right) \, \right) Br(H_2^0/A_2 \to \tau \bar{\tau}) > 25~{\rm fb},
\label{tautau} \\
&1.2~{\rm fb} > \left( \, \sigma_{{\rm 13 \, TeV}}\left(p p \to H_2^0\right)+\sigma_{{\rm 13 \, TeV}}\left(p p \to A_2\right) \, \right) Br(H_2^0/A_2 \to \gamma \gamma) > 0.74~{\rm fb},
\label{gammagamma} \\
&5.5~{\rm fb} > \sigma_{{\rm 13 \, TeV}}\left(p p \to H_2^+ H_2^-\right) Br(H_2^+ \to t \bar{b})^2 > 2.7~{\rm fb},
\nonumber \\
&10~{\rm fb} > \left( \, \sigma_{{\rm 13 \, TeV}}\left(p p \to H_2^0 H_2^\pm\right) + \sigma_{{\rm 13 \, TeV}}\left(p p \to A_2 H_2^\pm\right) \, \right) Br(H_2^+ \to t \bar{b})Br(H_2^0/A_2 \to b \bar{b}) > 3.9~{\rm fb},
\label{tb} \\
&1.3~{\rm fb} > \left( \, \sigma_{{\rm 13 \, TeV}}\left(p p \to H_2^0 H_2^\pm\right) + \sigma_{{\rm 13 \, TeV}}\left(p p \to A_2 H_2^\pm\right) \, \right) Br(H_2^+ \to t \bar{b})Br(H_2^0/A_2 \to \tau \bar{\tau}) > 0.54~{\rm fb},
\nonumber \\
&0.18~{\rm fb} > \sigma_{{\rm 13 \, TeV}}\left(p p \to H_2^0 A_2\right) Br(H_2^0/A_2 \to b \bar{b})Br(H_2^0/A_2 \to \tau \bar{\tau}) > 0.057~{\rm fb}.
\label{btau}
\end{align}
Here, the gluon fusion cross section is obtained by rescaling corresponding cross sections for a SM-like Higgs boson in Ref.~\cite{wg} by $(s_H/c_H)^2$,
 and the Drell-Yan cross sections are computed at tree level with CTEQ6L1~\cite{cteq} parton distribution function by MadGraph5aMC@NLO~\cite{mg}.
Eq.~(\ref{tautau}) is confronted with the search for a single heavy Higgs boson decaying into a $\tau$ pair with 13.3~fb$^{-1}$ of 13~TeV $pp$ collision data~\cite{heavyhiggstotautau},
 which reports the 95\% certainty level (CL) bound on the cross section times branching fraction to be 1~pb for $M_2=266$~GeV and 0.4~pb for $M_2=316$~GeV,
 and hence our model is not constrained.
Eq.~(\ref{gammagamma}) is confronted with the search for a diphoton resonance with 15.4~fb$^{-1}$ of 13~TeV $pp$ collision data~\cite{diphoton},
 which reports the 95\% CL bound on the cross section times branching fraction to be 10~fb for about 300~GeV invariant mass, and hence our model is not constrained.
The processes of Eq.~(\ref{tb}) mimic the event topology of
 "single charged Higgs boson production associated with $\bar{t}b$, followed by the decay into $t\bar{b}$"
 which is searched for with 13.2~fb$^{-1}$ of 13~TeV $pp$ collision data~\cite{ttbarbb}.
The 95\% CL bound on the cross section times branching fraction is 1~pb when a $t\bar{b}$ pair is not boosted, which is the case for our model,
 and hence the model safely evades the constraint.
For the processes of Eq.~(\ref{btau}), no corresponding report with 13~TeV collision data is found.

The model is constrained by the measurement of $b \to s \gamma$ decay, as $H_2^\pm$ particle contributes to the process.
The bound on $M_2$ in our model equals that on the charged Higgs boson mass in Type-I two Higgs doublet model with $\tan \beta = c_H/s_H$ in Ref.~\cite{bsgamma}.
Since $3.6>c_H/s_H>3.0$, our model is not excluded.

We discuss experimental implications of the $U(1)_X$ gauge boson.
Below the confinement scale, the $U(1)_X$ gauge boson couples exclusively to SM fermions and SM-gauge-singlet neutrinos,
 and the mass is given by $M_X = 2\sqrt{2}g_X \, f_\Pi$, with $770~{\rm GeV} < f_\Pi < 931~{\rm GeV}$ and $g_X$ denoting the $U(1)_X$ gauge coupling constant corresponding to the charge assignment in Table~\ref{content}.
This gauge boson is not experimentally ruled out if $10^{-5} \, e < g_X < 10^{-3} \, e$ ($e$ is the electric charge) and thus $10~{\rm MeV} < M_X < 1$~GeV, 
 where the lower bound guarantees that it is free from cosmological and astrophysical constraints.
\footnote{
Gauge bosons with such values of $g_X$ and $M_X$ are called "visible dark photon".
}
For $x=O(1)$, and unless $x\simeq0$ so that the couplings with leptons are non-negligible, 
 the most stringent upper bound on $g_X$ derives from the search for dark photon $A'$ in the process $e^+ e^- \to \gamma A', \, A' \to e^+ e^-,\mu^+ \mu^-$~\cite{babar}.
The reported bound is translated into the bound $x \, g_X \lesssim 5\times10^{-4} \, e$ in our model,
 but the actual bound is weaker because the $U(1)_X$ gauge boson can also decay into neutrinos.
We comment that a lower bound on $g_X$ cannot be obtained from the electron-beam-dump experiment searching for bremsstrahlung production of dark photon off an electron followed by the decay into $e^+ e^-$ reported in Ref.~\cite{dump},
 because $g_X$ is always smaller than the experiment's coverage due to the relation $M_X = 2\sqrt{2}g_X \, f_\Pi$.
For $x\simeq0$, the $U(1)_X$ gauge boson is produced from $q\bar{q}$, and if kinematically allowed, by rare meson decays~\cite{raremesondecays},
 and decays dominantly into $\pi^+ \pi^-$ when $1~{\rm GeV} \geq M_X \gtrsim 0.3$~GeV.
In the special case when $x=0$ and $M_X < 0.28$~GeV, the $U(1)_X$ gauge boson decays into three photons through its vectorial coupling with quarks,
 analogously to the ortho-positronium decay into three photons.
There is no bound for the $U(1)_X$ gauge boson in the above two cases where it decays into $\pi^+\pi^-$ or into three photons.
\\

We have presented an extension of the Standard Model where the multiple-point principle is realized.
Our model bears classical scale invariance, and the Standard Model Higgs field with a tachyonic mass emerges through the mixing of an elementary massless scalar field $H$ and a pseudo-Nambu-Goldstone field $\Pi$ coming from $SU(2)_T$ strongly-coupled gauge theory.
The Standard Model Higgs quartic coupling is induced from that for $H$,
 which gives a gap between the two quartic couplings and
 allows the $H$ quartic coupling to satisfy the conditions for the multiple-point principle without contradicting the measured Higgs particle mass.
By solving the renormalization group equations, we have derived the pseudo-Nambu-Goldstone boson mass and the top quark pole mass with which the multiple-point principle is attained.
Based on the pseudo-Nambu-Goldstone boson mass thus obtained, we have predicted the mass spectrum of new particles and their experimental signatures.
\\

\section*{Acknowledgement}

This work is partially supported by Scientific Grants by the Ministry of Education, Culture, Sports, Science and Technology of Japan (Nos. 24540272, 26247038, 15H01037, 16H00871, and 16H02189).
\\


\begin{thebibliography}{99}
 \bibitem{mpp}
   C.~D.~Froggatt and H.~B.~Nielsen,
  ``Standard model criticality prediction: Top mass 173 +- 5-GeV and Higgs mass 135 +- 9-GeV,''
  Phys.\ Lett.\ B {\bf 368}, 96 (1996)
  [hep-ph/9511371].
 
 \bibitem{csi}
   W.~A.~Bardeen,
  ``On naturalness in the standard model,''
  FERMILAB-CONF-95-391-T, C95-08-27.3.

\bibitem{combinedhiggs}
   G.~Aad {\it et al.} [ATLAS and CMS Collaborations],
  ``Combined Measurement of the Higgs Boson Mass in $pp$ Collisions at $\sqrt{s}=7$ and 8 TeV with the ATLAS and CMS Experiments,''
  Phys.\ Rev.\ Lett.\  {\bf 114}, 191803 (2015)
  [arXiv:1503.07589 [hep-ex]].

\bibitem{rge}
  D.~Buttazzo, G.~Degrassi, P.~P.~Giardino, G.~F.~Giudice, F.~Sala, A.~Salvio and A.~Strumia,
  ``Investigating the near-criticality of the Higgs boson,''
  JHEP {\bf 1312}, 089 (2013)
  [arXiv:1307.3536 [hep-ph]];

\bibitem{rgeanalyses}
  C.~D.~Froggatt, H.~B.~Nielsen and Y.~Takanishi,
  ``Standard model Higgs boson mass from borderline metastability of the vacuum,''
  Phys.\ Rev.\ D {\bf 64}, 113014 (2001)
  [hep-ph/0104161];
  H.~B.~Nielsen,
  ``PREdicted the Higgs Mass,''
  Bled Workshops Phys.\  {\bf 13}, no. 2, 94 (2012)
  [arXiv:1212.5716 [hep-ph]];
  M.~Holthausen, K.~S.~Lim and M.~Lindner,
  ``Planck scale Boundary Conditions and the Higgs Mass,''
  JHEP {\bf 1202}, 037 (2012)
  [arXiv:1112.2415 [hep-ph]];
  F.~Bezrukov, M.~Y.~Kalmykov, B.~A.~Kniehl and M.~Shaposhnikov,
  ``Higgs Boson Mass and New Physics,''
  JHEP {\bf 1210}, 140 (2012)
  [arXiv:1205.2893 [hep-ph]];
  G.~Degrassi, S.~Di Vita, J.~Elias-Miro, J.~R.~Espinosa, G.~F.~Giudice, G.~Isidori and A.~Strumia,
  ``Higgs mass and vacuum stability in the Standard Model at NNLO,''
  JHEP {\bf 1208}, 098 (2012)
  [arXiv:1205.6497 [hep-ph]];
  S.~Alekhin, A.~Djouadi and S.~Moch,
  ``The top quark and Higgs boson masses and the stability of the electroweak vacuum,''
  Phys.\ Lett.\ B {\bf 716}, 214 (2012)
  [arXiv:1207.0980 [hep-ph]];
  I.~Masina,
  ``Higgs boson and top quark masses as tests of electroweak vacuum stability,''
  Phys.\ Rev.\ D {\bf 87}, no. 5, 053001 (2013)
  [arXiv:1209.0393 [hep-ph]];
  Y.~Hamada, H.~Kawai and K.~y.~Oda,
  ``Bare Higgs mass at Planck scale,''
  Phys.\ Rev.\ D {\bf 87}, no. 5, 053009 (2013)
  Erratum: [Phys.\ Rev.\ D {\bf 89}, no. 5, 059901 (2014)]
  [arXiv:1210.2538 [hep-ph]];
  F.~Jegerlehner,
  ``The Standard model as a low-energy effective theory: what is triggering the Higgs mechanism?,''
  Acta Phys.\ Polon.\ B {\bf 45}, no. 6, 1167 (2014)
  [arXiv:1304.7813 [hep-ph]];
  I.~Masina and M.~Quiros,
  ``On the Veltman Condition, the Hierarchy Problem and High-Scale Supersymmetry,''
  Phys.\ Rev.\ D {\bf 88}, 093003 (2013)
  [arXiv:1308.1242 [hep-ph]];
  A.~Spencer-Smith,
  ``Higgs Vacuum Stability in a Mass-Dependent Renormalisation Scheme,''
  arXiv:1405.1975 [hep-ph].
  
 \bibitem{mppwithmass}
   N.~Haba, H.~Ishida, N.~Okada and Y.~Yamaguchi,
  ``Multiple-point principle with a scalar singlet extension of the Standard Model,''
  PTEP {\bf 2017}, no. 1, 013B03 (2017)
  [arXiv:1608.00087 [hep-ph]].
  
 \bibitem{bosonicseesaw}
   X.~Calmet,
  ``Seesaw induced Higgs mechanism,''
  Eur.\ Phys.\ J.\ C {\bf 28}, 451 (2003)
  [hep-ph/0206091];
   H.~D.~Kim,
  ``Electroweak symmetry breaking from SUSY breaking with bosonic see-saw mechanism,''
  Phys.\ Rev.\ D {\bf 72}, 055015 (2005)
  [hep-ph/0501059];
   N.~Haba, N.~Kitazawa and N.~Okada,
  ``Invisible technicolor,''
  Acta Phys.\ Polon.\ B {\bf 40}, 67 (2009)
  [hep-ph/0504279].
  
  \bibitem{similar}
     O.~Antipin, M.~Redi and A.~Strumia,
  ``Dynamical generation of the weak and Dark Matter scales from strong interactions,''
  JHEP {\bf 1501}, 157 (2015)
  [arXiv:1410.1817 [hep-ph]];
    K.~Kannike, G.~M.~Pelaggi, A.~Salvio and A.~Strumia,
  ``The Higgs of the Higgs and the diphoton channel,''
  JHEP {\bf 1607}, 101 (2016)
  [arXiv:1605.08681 [hep-ph]];
    N.~Haba, H.~Ishida, N.~Kitazawa and Y.~Yamaguchi,
  ``A new dynamics of electroweak symmetry breaking with classically scale invariance,''
  Phys.\ Lett.\ B {\bf 755}, 439 (2016)
  [arXiv:1512.05061 [hep-ph]];
     H.~Ishida, S.~Matsuzaki and Y.~Yamaguchi,
  ``Invisible Axion-Like Dark Matter from Electroweak Bosonic Seesaw,''
  arXiv:1604.07712 [hep-ph];
    H.~Ishida, S.~Matsuzaki and Y.~Yamaguchi,
  ``Bosonic-Seesaw Portal Dark Matter,''
  arXiv:1610.07137 [hep-ph];
    H.~Ishida, S.~Matsuzaki, S.~Okawa and Y.~Omura,
  ``Scalegenesis via dynamically induced multiple seesaws,''
  arXiv:1701.00598 [hep-ph];
    N.~Haba and T.~Yamada,
  ``Strong dynamics in a classically scale invariant extension of the Standard Model with flatland,''
  arXiv:1701.02146 [hep-ph].




\bibitem{wzw}
    J.~Wess and B.~Zumino,
  ``Consequences of anomalous Ward identities,''
  Phys.\ Lett.\  {\bf 37B}, 95 (1971);
  E.~Witten,
  ``Global Aspects of Current Algebra,''
  Nucl.\ Phys.\ B {\bf 223}, 422 (1983).
 
  

 \bibitem{tc}
   S.~Weinberg,
  ``Implications of Dynamical Symmetry Breaking,''
  Phys.\ Rev.\ D {\bf 13}, 974 (1976);
    L.~Susskind,
  ``Dynamics of Spontaneous Symmetry Breaking in the Weinberg-Salam Theory,''
  Phys.\ Rev.\ D {\bf 20}, 2619 (1979).







\bibitem{mac}
    S.~Raby, S.~Dimopoulos and L.~Susskind,
  ``Tumbling Gauge Theories,''
  Nucl.\ Phys.\ B {\bf 169}, 373 (1980).



 
 \bibitem{dashen}
  R.~F.~Dashen,
  ``Chiral SU(3) x SU(3) as a symmetry of the strong interactions,''
  Phys.\ Rev.\  {\bf 183}, 1245 (1969).
 
 
 \bibitem{scalarmesons}
  R.~L.~Jaffe,
  ``Multi-Quark Hadrons. 1. The Phenomenology of (2 Quark 2 anti-Quark) Mesons,''
  Phys.\ Rev.\ D {\bf 15}, 267 (1977);
  R.~L.~Jaffe,
  ``Multi-Quark Hadrons. 2. Methods,''
  Phys.\ Rev.\ D {\bf 15}, 281 (1977);
  F.~E.~Close and N.~A.~Tornqvist,
  ``Scalar mesons above and below 1-GeV,''
  J.\ Phys.\ G {\bf 28}, R249 (2002)
  [hep-ph/0204205];
    R.~L.~Jaffe and F.~Wilczek,
  ``Diquarks and exotic spectroscopy,''
  Phys.\ Rev.\ Lett.\  {\bf 91}, 232003 (2003)
  [hep-ph/0307341];
  L.~Maiani, F.~Piccinini, A.~D.~Polosa and V.~Riquer,
  ``A New look at scalar mesons,''
  Phys.\ Rev.\ Lett.\  {\bf 93}, 212002 (2004)
  [hep-ph/0407017];
   G.~'t Hooft, G.~Isidori, L.~Maiani, A.~D.~Polosa and V.~Riquer,
  ``A Theory of Scalar Mesons,''
  Phys.\ Lett.\ B {\bf 662}, 424 (2008)
  [arXiv:0801.2288 [hep-ph]].
 
 
 \bibitem{pdg}
 C.~Patrignani \textit{et al.} (Particle Data Group), Chin.\ Phys.\ C {\bf 40}, 100001 (2016). 
 
 \bibitem{durr}
  S.~Borsanyi, S.~Durr, Z.~Fodor, S.~Krieg, A.~Schafer, E.~E.~Scholz and K.~K.~Szabo,
  ``SU(2) chiral perturbation theory low-energy constants from 2+1 flavor staggered lattice simulations,''
  Phys.\ Rev.\ D {\bf 88}, 014513 (2013)
  [arXiv:1205.0788 [hep-lat]].

 
 \bibitem{atlastoppole}
 G.~Aad {\it et al.} [ATLAS Collaboration],
  ``Determination of the top-quark pole mass using $ t\overline{t} $ + 1-jet events collected with the ATLAS experiment in 7 TeV pp collisions,''
  JHEP {\bf 1510}, 121 (2015)
  [arXiv:1507.01769 [hep-ex]].
 
 
 \bibitem{atlastop}
   M.~Aaboud {\it et al.} [ATLAS Collaboration],
  ``Measurement of the top quark mass in the $t\bar{t}\to$ dilepton channel from $\sqrt{s}=8$ TeV ATLAS data,''
  Phys.\ Lett.\ B {\bf 761}, 350 (2016)
  [arXiv:1606.02179 [hep-ex]].

 \bibitem{cmstop}
   V.~Khachatryan {\it et al.} [CMS Collaboration],
  ``Measurement of the top quark mass using proton-proton data at ${\sqrt{(s)}}$ = 7 and 8 TeV,''
  Phys.\ Rev.\ D {\bf 93}, no. 7, 072004 (2016)
  [arXiv:1509.04044 [hep-ex]].
 
 
 \bibitem{polevsmc}
   A.~H.~Hoang and I.~W.~Stewart,
  ``Top Mass Measurements from Jets and the Tevatron Top-Quark Mass,''
  Nucl.\ Phys.\ Proc.\ Suppl.\  {\bf 185}, 220 (2008)
  doi:10.1016/j.nuclphysbps.2008.10.028
  [arXiv:0808.0222 [hep-ph]];
  A.~H.~Hoang, A.~Jain, I.~Scimemi and I.~W.~Stewart,
  ``Infrared Renormalization Group Flow for Heavy Quark Masses,''
  Phys.\ Rev.\ Lett.\  {\bf 101}, 151602 (2008)
  [arXiv:0803.4214 [hep-ph]];
   M.~Butenschoen, B.~Dehnadi, A.~H.~Hoang, V.~Mateu, M.~Preisser and I.~W.~Stewart,
  ``Top Quark Mass Calibration for Monte Carlo Event Generators,''
  Phys.\ Rev.\ Lett.\  {\bf 117}, no. 23, 232001 (2016)
  [arXiv:1608.01318 [hep-ph]].


\bibitem{wg}
  S.~Heinemeyer {\it et al.} [LHC Higgs Cross Section Working Group],
  ``Handbook of LHC Higgs Cross Sections: 3. Higgs Properties,''
  arXiv:1307.1347 [hep-ph].


\bibitem{cteq}
  J.~Pumplin, D.~R.~Stump, J.~Huston, H.~L.~Lai, P.~M.~Nadolsky and W.~K.~Tung,
  ``New generation of parton distributions with uncertainties from global QCD analysis,''
  JHEP {\bf 0207}, 012 (2002)
  [hep-ph/0201195].


\bibitem{mg}
  J.~Alwall {\it et al.},
  ``The automated computation of tree-level and next-to-leading order differential cross sections, and their matching to parton shower simulations,''
  JHEP {\bf 1407}, 079 (2014)
  [arXiv:1405.0301 [hep-ph]].

\bibitem{heavyhiggstotautau}
https://atlas.web.cern.ch/Atlas/GROUPS/PHYSICS/CONFNOTES/ATLAS-CONF-2016-085/

\bibitem{diphoton}
https://atlas.web.cern.ch/Atlas/GROUPS/PHYSICS/CONFNOTES/ATLAS-CONF-2016-059/

\bibitem{ttbarbb}
https://atlas.web.cern.ch/Atlas/GROUPS/PHYSICS/CONFNOTES/ATLAS-CONF-2016-104/

\bibitem{bsgamma}
  M.~Misiak and M.~Steinhauser,
  ``Weak Radiative Decays of the B Meson and Bounds on $M_{H^\pm}$ in the Two-Higgs-Doublet Model,''
  arXiv:1702.04571 [hep-ph].

\bibitem{babar}
  J.~P.~Lees {\it et al.} [BaBar Collaboration],
  ``Search for a Dark Photon in $e^+e^-$ Collisions at BaBar,''
  Phys.\ Rev.\ Lett.\  {\bf 113}, no. 20, 201801 (2014)
  [arXiv:1406.2980 [hep-ex]].

\bibitem{dump}
  E.~M.~Riordan {\it et al.},
  ``A Search for Short Lived Axions in an Electron Beam Dump Experiment,''
  Phys.\ Rev.\ Lett.\  {\bf 59}, 755 (1987).

\bibitem{raremesondecays}
  M.~Reece and L.~T.~Wang,
  ``Searching for the light dark gauge boson in GeV-scale experiments,''
  JHEP {\bf 0907}, 051 (2009)
  [arXiv:0904.1743 [hep-ph]].


\end{thebibliography}
\end{document}